Comment on: High Mixing Entropy Enhanced Energy States in Metallic Glasses

Ramir Ristić[1] and Emil Babić[2]

[1]Department of Physics, University of Osijek, Trg Ljudevita Gaja 6, HR-31000 Osijek, Croatia

[2]Department of Physics, Faculty of Science, Bijenička cesta 32, HR-10002 Zagreb, Croatia

A recent paper by Juntao Huo et al [1] reported a correlation between the entropy of mixing, $\Delta S_{mix}$ (and the corresponding energy state) and the thermal stability and mechanical parameters, for three Zr-Ti-Cu-Ni-Be metallic glasses (MG) including a high-entropy one (HEMG). Namely a monotonic increase in the thermal stability and strength with $\Delta S_{mix}$ was observed. Although the enhancement of thermal stability of HEMGs by $\Delta S_{mix}$ has already been proposed [2,3] (see [4] for an alternative explanation) and the correlation between the thermal stability and mechanical parameters of MGs has been well established [4-7], the correlation reported in [1] is potentially important and deserves further experimental verification.

The authors in [1] dismissed a possible compositional contribution to the variations of the thermal stability and mechanical parameters studied (Fig. 1 b and 4, Table 1 of [1]) which as illustrated in Fig. 1 below may not be justified. As seen in Fig. 1 the variation of the crystallization temperature $T_x$ and of the yield strength $\sigma_y$ with the total amount of late transition metals (TL), $x=x_{Cu}+x_{Ni}$ (in atomic percent) is qualitatively the same as the variation with $\Delta S_{mix}$ in Fig. 1 b and 4 a in [1]. The same is true for all other parameters studied in [1]. The increase of the thermal stability and mechanical parameters of MGs combining early transition metals (TE= Ti, Zr and Hf) and late ones (TL=Co, Ni and Cu), with increasing TL content is well known and reflects the enhancement of interatomic bonding on alloying (amorphous) TEs with TLs [4,7-9]. This enhancement is specific to TE-TL MGs and does not depend on the number of components in a given alloy [4, 10-12]. So it is unlikely to be caused only by changes in the entropy of mixing.

A study of a new $Ti_{40}Zr_{12}Cu_{25}Ni_3Be_{20}$ MG may provide a simple way of determining the contributions of x and $\Delta S_{mix}$ to the thermal stability and strength of TE-TL MGs since this alloy has the same $\Delta S_{mix}$ as the alloy Ti40 in [1], while having considerably larger x=28 compared to that of Ti40 (x=15).

Acknowledgements. We thank Professor J.R. Cooper for critical reading of this Comment.

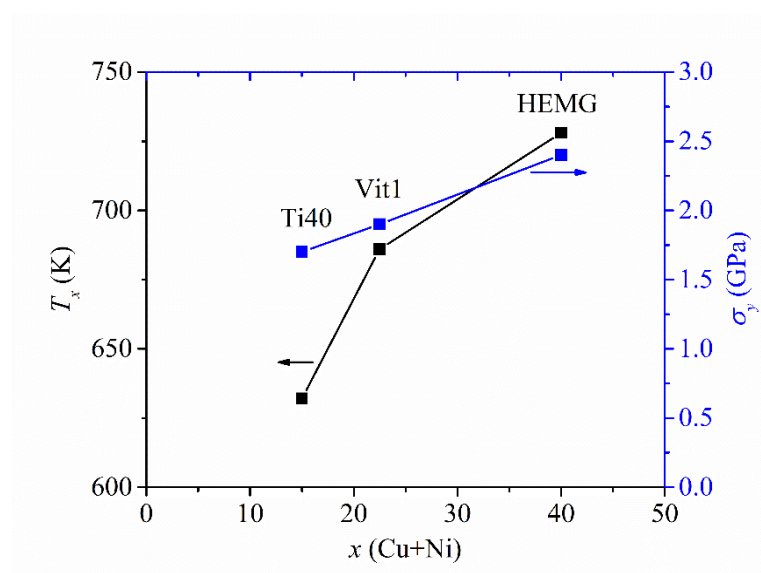

Fig.1 The variation of initial crystallization temperature $T_x$ and yield strength $\sigma_y$ of $Zr_{20}Ti_{20}Cu_{20}Ni_{20}Be_{20}$, $Zr_{41}Ti_{14}Cu_{12.5}Ni_{10}Be_{22.5}$ and $Ti_{40}Zr_{25}Cu_{12}Ni_3Be_{20}$ (Ti40) metallic glasses[1] with total content of Cu and Ni, $x=x_{Cu}+x_{Ni}$.